\documentclass[sigconf,natbib=true,anonymous=false]{acmart}
\AtBeginDocument{%
  \providecommand\BibTeX{{%
    \normalfont B\kern-0.5em{\scshape i\kern-0.25em b}\kern-0.8em\TeX}}}
\usepackage{multirow}
\usepackage{amsmath,amsthm,mathtools}
\usepackage{subfigure} 
\title{
Multi-Granularity Attention Model for Group Recommendation
}

\author{Jianye Ji\textsuperscript{*}, Jiayan Pei\textsuperscript{*}, Shaochuan Lin\textsuperscript{*}, Taotao Zhou, Hengxu He, Jia Jia, Ning Hu}
\affiliation{%
  \institution{Alibaba Group}
  \city{Hangzhou\&Shanghai}
  \country{China}
}
\email{{jianyeji.jjy, jiayanpei.pjy, lin.lsc, taotao.zhou, hengxu.hhx, jj229618, huning.hu}@alibaba-inc.com}

\thanks{* These authors contributed equally}

\date{October 2023}
\acmConference[Conference'23]{}{October 2023}{Birmingham, UK}

\begin{document}

\begin{abstract}

% either heavily rely on the potential preferences of users with rich behavior or ignore the potential preferences of users with sparse behavior, resulting in insufficient learning of individual interests.

% most of them either overlook the impact of individuals or excessively focus on it, resulting in a disregard for distinct preferences of group members or an excess of recommendation noise.

 % enhance individual preferences and minimize recommendation noise.

% gather and aggregate information between users and items. 

% XXX1 and XXX2 propose to further investigate users' latent preferences on the group-level(which maintains users original preference) and superset-level (which includes group-group exterior information), respectively.
Group recommendation provides personalized recommendations to a group of users based on their shared interests, preferences, and characteristics. 
Current studies have explored different methods for integrating individual preferences and making collective decisions that benefit the group as a whole. However, most of them heavily rely on users with rich behavior and ignore latent preferences of users with relatively sparse behavior, leading to insufficient learning of individual interests. To address this challenge, we present the Multi-Granularity Attention Model (MGAM), a novel approach that utilizes multiple levels of granularity ($i.e.$, subsets, groups, and supersets) to uncover group members' latent preferences and mitigate recommendation noise.
Specially, we propose a Subset Preference Extraction module that enhances the representation of users' latent subset-level preferences by incorporating their previous interactions with items and utilizing a hierarchical mechanism. Additionally, our method introduces a Group Preference Extraction module and a Superset Preference Extraction module, which explore users' latent preferences on two levels: the group-level, which maintains users' original preferences, and the superset-level, which includes group-group exterior information. By incorporating the subset-level embedding, group-level embedding, and superset-level embedding, our proposed method effectively reduces group recommendation noise across multiple granularities and comprehensively learns individual interests.
Extensive offline and online experiments have demonstrated the superiority of our method in terms of performance.

\end{abstract}

\maketitle

%\begin{keywords}
%kkeywords: Group %recommendation, %hierarchical attention, %%\end{keywords}

\section{Introduction}
% In the information overloading world, recommender systems are widely employed owing to their ability to effectively connect users and their highly desired information\cite{huang2020efficient}. However, 

% With the rapid prevalence of social media and online communities, people prefer to organize and participate in group activities \cite{deng2016deep, huang2016pairwise}, such as group trips, family dinners and colleagues parties. Therefore, it is an urgent trend for recommend systems to provide appropriate recommendations for group activities, referred to as the \textit{group recommendation task}.

% \textcolor{red}{talk about the importance of group recommendation, especially compared to individual recommendation. }

Group recommendation offers tailored and attractive suggestions to each member of the group, ensuring personalized satisfaction for all. Compared with individual recommendations, group recommendations provide a broader range of options by considering social influence and collaborative decision-making, making them especially beneficial for group activities \cite{huang2020efficient, SIGR} such as family dinners, group trips, and colleagues' parties.

In recent research, group recommendation has garnered significant attention due to its widespread application. Prior studies \cite{baltrunas2010group, amer2009group, yu2006tv, boratto2014modeling} have employed static strategies to aggregate users' profiles, resulting in limited consideration of diverse member preferences. To address this gap, neural attention-based methods \cite{AGREE, cao2019social,   vinh2019interact, sankar2020groupim, huang2020efficient, wang2020group, liu2019nrpa, 2019GRADI} have been proposed to dynamically adjust members' weights. However, these methods often suffer from overfitting and noise due to their heavy reliance on users with rich behavioral data. Moreover, assigning lower weights to users with fewer interactions may fail to capture their preferences, even if they have a rich history of interactions with other items. Graph-based methods \cite{GAME, SIGR, HHE,  HANCDGR, ConsRec, S2HHGR} have attempted to integrate intra- and inter-graph connections to alleviate this issue, but they still lack the ability to utilize comprehensive granularity for better representation of the final group embedding in multi-views.

% can be categoried into two approaches,  memory-based methods and model-based methods. 

% Individual: ARGREE \cite{AGREE}, 

% Graph: GAME \cite{GAME}, ConsRec \cite{ConsRec}, HHE \cite{HHE}, HANCDGR \cite{HANCDGR}, SIGR \cite{SIGR}, S2HHGR \cite{S2HHGR} intra-inter feature enhancing 

% Group decision-making is a challenging task as each member may have varying preferences, which can significantly impact the final recommendation \cite{SIGR}. Researches over group recommendations mainly focus on learning each member’s preferences over items and taking the appropriate strategy to aggregate them to make the decision, which can be generally divided into memory-based and model-based approaches. Memory-based methods aggregate users’ profiles through pre-defined strategies \cite{baltrunas2010group, amer2009group, yu2006tv, boratto2014modeling}, while model-based approaches generate the group decision-making process to make recommendations in a learnable way \cite{liu2012exploring, yuan2014generative, hu2014deep}. 

% However, due to the defect that the
% memory-based approaches are limited in their ability to consider diverse member preferences, model-based approaches integrate attention mechanisms \cite{AGREE, cao2019social, SIGR,  vinh2019interact, sankar2020groupim, huang2020efficient, wang2020group, liu2019nrpa, 2019GRADI, HHE, GAME, HANCDGR, ConsRec}, which further adjust the weights of members dynamically, have been the practical methods in current group recommendation systems.
% \textcolor{red}{Scale into One Paragraph with the above.}

\begin{figure}[tbp]
  \centering
  % \vspace{5pt}
  \includegraphics[width=0.8\linewidth]{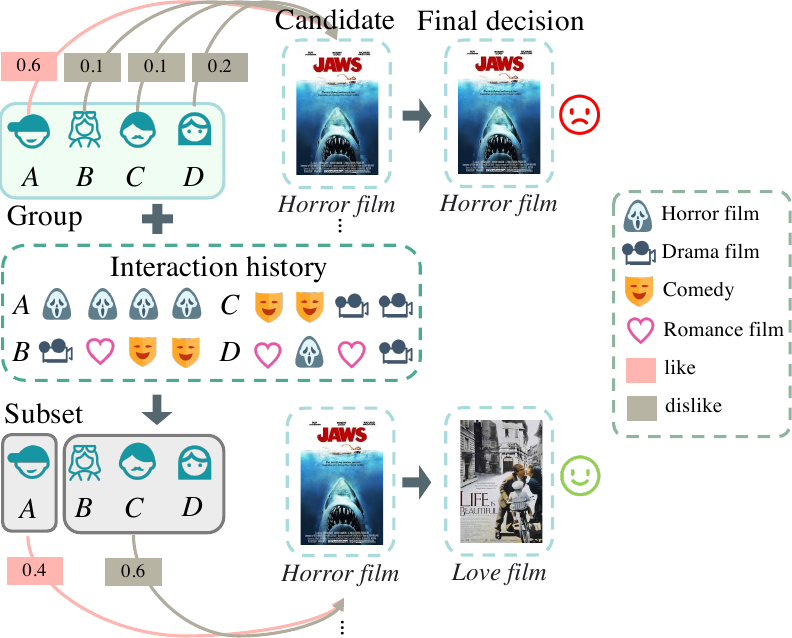}
  \vspace{-10pt}
  \caption{Example showcasing the significance of subset granularity. User A's interactions with a candidate item carry significant weight compared to users with less interaction, but without considering the granularity of subsets. By utilizing the interaction history with other items to incorporate subset level granularity, we can arrive at a more informed and acceptable final decision for the entire group.}
  \vspace{-15pt}
  \label{fig:intro}
\end{figure}

In this paper, we propose a solution to the above challenge of uncovering group members' latent preferences and mitigating recommendation noise through the \textbf{M}ulti-\textbf{G}ranularity \textbf{A}ttention \textbf{M}odel (\textbf{MGAM}). The MGAM effectively leverages multi-granularity at three levels: subset, group, and superset, to achieve this goal. Specifically, we introduce the Subset Preference Extraction (SubPE) module to incorporate a new level of granularity, namely subset, to improve the representation of users' latent subset-level preferences. The SubPE module leverages the users' past interactions with other items and employs a hierarchical mechanism to mitigate the reliance on users with abundant behavioral data on candidate items. Fig.~\ref{fig:intro} illustrates the role that subset-level granularity plays in our proposed method. Additionally, our approach involves utilizing the Group Preference Extraction (GPE) module to create a unique group-level preference representation that preserves initial user expression and aids decision-making. To reduce recommendation noise and gain a deeper understanding of group representation, we employ the Superset Preference Extraction (SupPE) module at the superset-level granularity to capture external preference information. Our fusion layer dynamically incorporates subset-level, group-level, and superset-level embeddings to uncover latent but real preferences of group members. Extensive experiments on public datasets demonstrate the superiority of our approach over state-of-the-art methods, and successful online A/B testing further proves its effectiveness.

\section{Methodology}
\begin{figure}[tbp]
  \centering
  \vspace{-10pt}
  \includegraphics[width=0.95\linewidth]{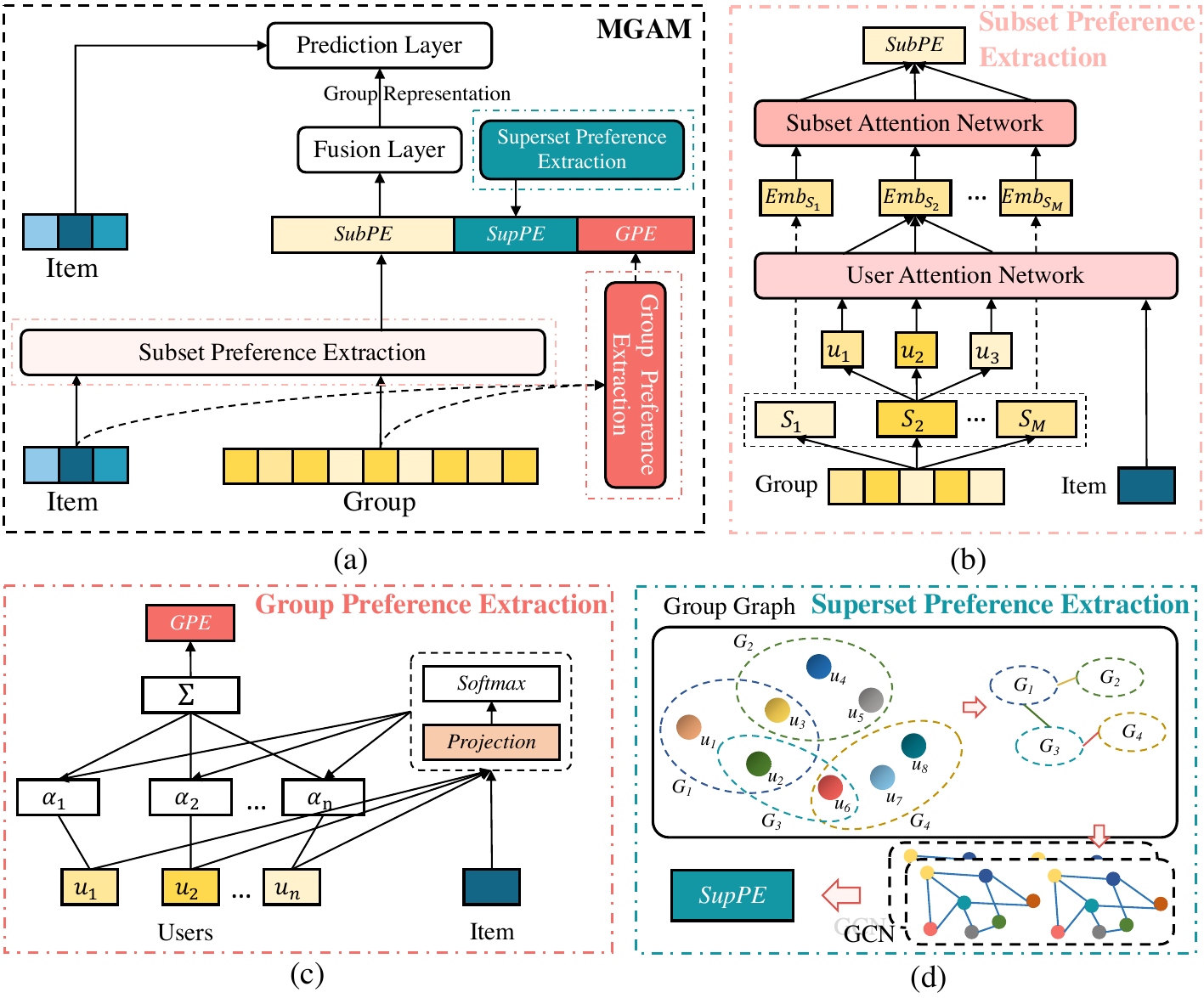}
  \vspace{-10pt}
  \caption{The overview of MGAM (a) and each substructure for preference extraction (b, c, d).}
  \vspace{-20pt}
  \label{fig: model struct}
\end{figure}

%For group recommendation, the main task is to calculate the preference aggregation based on a group profile that combines all user preferences. So our method mainly focuses on group user aggregation.
% Group recommendation aims to make the decision based on the group profile aggregating from all members' preferences. It is essential to take the appropriate strategy for the preferences' aggregation. Also, information outside the group brings a marginal contribution. Figure 2 gives an illustration of the GHAN framework. 
In this section, we introduce the definition of group recommendation task and detail the MGAM, with its overall architecture illustrated in Fig.~\ref{fig: model struct}. MGAM proposes the Subset Preference Extraction Module, the Group Preference Extraction Module, and the Superset Preference Extraction Module, which comprehensively solves the above problems from the subset granularity, the group granularity, and the superset granularity, respectively.

\subsection{Preliminaries}
In group recommendation, there are three sets of entities: a user set $\mathcal{U} = \{u_1, u_2, ..., u_n\}$, an item set $\mathcal{V} = \{v_1, v_2, ..., v_p\}$, and a group set $\mathcal{G} = \{g_1, g_2, ..., g_q\}$, where $n$, $p$, $q$ denote the sizes of these three sets. The $t$-th group $g_t \in \mathcal{G}$ consists of a set of user members $g_t = \{u_1, u_2, ..., u_{\lvert gl \vert}\}$, where $u_{\lvert gl \vert} \in \mathcal{U}$ and $\lvert g_l \vert$ is the size of $g_t$. The aim of this task is to calculate the probability of an item being interacted with by a specific group $g_t$,
\begin{equation}
    \mathcal{P}(y_i=1|x_i) = \mathcal{F}_{\theta}(v_i, g_t, \mathcal{I}_{u_i}, \mathcal{I}_{g_t}) 
\end{equation}
where the training data is represented as $\{{x_i, y_i}\}_{i=1}^{N}\in \mathcal{D}$, $i$ refers to the index of the data and $N$ is the total number of data. $\mathcal{F}$ represents the group recommendation model with training parameter $\theta$. The label $y_i \in \mathcal{Y}$ indicates whether group $g_t$ has interacted with item $v_i$. $x_i \in \mathcal{X}$ is a sample consisting of $v_i$, $g_t$, $\mathcal{I}{u_i}$, and $\mathcal{I}_{g_t}$, representing item, group, and user-item and group-item interactions respectively.

\subsection{Subset Preference Extraction (SubPE)}
Our proposal introduces a new level of granularity called subsets, which improves the representation of users' preferences. We achieve this using a hierarchical mechanism consisting of the User Attention Network and Subset Attention Network, which reduces reliance on users with extensive behavioral data on candidate items. Further details can be found in Fig.~\ref{fig: model struct}(b).
\subsubsection{Subset Generation}
\label{sec:subset generation}
 
% Firstly, we employ the k-means algorithm \cite{macqueen1967some} into cluster the members within each group. Specifically, we cluster members to $M$ sub-groups based on the user-item interaction history. In this way, users within the same sub-group have similar characteristics and preferences.

% \textcolor{red}{talk about the weak of lacking subset level}
% Traditional group-granularity-based methods have issues of overfitting and noise due to heavy reliance on users with substantial behavioral data. Lower weights assigned to users with fewer interactions may not accurately depict their preferences. To solve this, subsets are introduced as a new level of granularity. Specially, we use a clustering method to group users within the existing group into $M$ subsets based on their interaction history, which is considered the most effective approach for understanding a user's preferences. This allows us to identify users with similar preferences and group them together to filter out any noise or bias in the decision-making process, resulting in a more robust and accurate understanding of user preferences. 
Traditional group-based methods often suffer from overfitting and noise, as they heavily rely on users with substantial behavioral data. This can result in an inaccurate depiction of preferences for users with fewer interactions. To address this, we introduce subsets as a new level of granularity, clustering users within existing groups based on their interaction history into $M$ subsets. This enables us to identify users with similar preferences and filter out noise and bias in decision-making, leading to a more robust and accurate understanding of user preferences. Thus, $g_t = \{ s_1, ..., s_i, ..., s_M\}$, where $s_i$ represents a specific subset.

% We use a clustering method to group users into subsets based on their interaction history, since the best way to understand a user's preference is by analyzing their interaction history,  allowing us to identify users with similar preferences and group them together. By introducing this intermediate level of granularity, subset, we can filter out noise or bias in the decision-making process and fully consider the unique preferences of each member and subset within the group. This approach recognizes the individualities within the commonalities and provides a more accurate understanding of user preferences. After conducting several experiments and comparisons, we have selected the k-means algorithm as the most effective method for generating subsets. The number of subsets $M$ is treated as a hyper-parameter and adjusted during tuning. \textcolor{red}{add some math abstract description here. }
% Additionally, by adopting a larger granularity than the individual to refine the modeling process of group recommendation decisions, we aim to address the issue of decision bias caused by the dominance of interactive and rich users in the decision-making results

% In our experiments, we set the  $M$ equal to 5 according to the elbow method. 

\subsubsection{Hierarchical Mechanism}
%Let $u_i$ be the user i embedding vector,$v_j$ be the item j embedding vector,$g_k,j$ be the group k embedding which comes from users aggregation that show preference on item j.$\hat{g}_{k,m,j}$  be the subgroup m of group k embedding vector.

To improve the representation of users' preferences, we employ a hierarchical mechanism that begins by consolidating similar preferences within each subset utilizing a User Attention Network. This is followed by a Subset Attention Network that reduces reliance on users with extensive behavioral data and improves weights for users with focused preferences. 

\textit{User Attention Network} aggregates preference within each subset:
\begin{equation}
h_s^i = \sum_{k=1}^m{\frac{exp(\phi(\mathbf{W}_u(e(u_k)^T \cdot e(v_j))+\mathbf{b}_u))}{\sum_{k \in m}{exp(\phi(\mathbf{W}_u(e(u_k)^T \cdot e(v_j))+\mathbf{b}_u))}} u_k} \\
% h_s = \sum_{i=1}^mSoftmax(\phi(\mathbf{W}_u(e(u_k)^T \cdot e(v_j))+b_u)) u_k
\end{equation}
where $e(u_k)$ and $e(v_j)$ represent the embeddings of $u_k$ and $v_j$, respectively. $j$ is the index of an item. $m$ is the size of subset $s_i$, $u_k \in s_i$, $k$ also refers to the index. $\mathbf{W}_u$ and $\mathbf{b}_u$ are trainable parameters, and $\phi$ is a ReLU activation function. $h_s^i$ is the preference embedding for subset $s_i$.

% In the User Aggregation Module, we use an attention network to dynamically learn users' weight within a sub-group to aggregate the sub-group preference,
% We convert the user embedding and item embedding to the hidden layer vector through a single layer network with a $ReLU$ activation function. Then the $softmax$ function is applied to normalize the users' scores to get the attention weight of each user within the sub-group. Finally, we can get the sub-group  embedding by weighted-sum:

% \begin{equation}
% \alpha_{k,m,i,j}=\frac{exp(\sigma(W_1(u_k^T \cdot v_j)+b_1))}{\sum_{i \in I}{exp(\sigma(W_1(u_k^T \cdot v_j)+b_1))}} \ \ \ \forall i \in [1,...,I] 
% \end{equation}

% where $e(u_k)$ and $e(v_j)$ are the embedding of $u_k$ and $v_j$, respectively. $j$ is the index of item. $u_k \in s_i$, $s_i$ has a size of $m$. $\mathbf{W}_u$ and $\mathbf{b}_u$ are learnable parameters, $\phi$ denotes the ReLU activation. $h_s$ refers to subset $s_i$ aggregation preference embedding.
% \textcolor{red}{where ... }

\textit{Subset Attention Network} uses an attention mechanism like MoSAN \cite{vinh2019interact} to capture interactions and variations between subsets, giving more weight to users with specific preferences and improving overall performance. This can be visually represented in Fig.~\ref{fig:intro}. Finally, we obtain the aggregated SubPE embedding for group $g_t$, 
\begin{gather}
    a_i=\mathbf{W}_s\phi(\mathbf{W}_{i}{h}_{s}^i +\overline{\mathbf{W}}_{i}\bar{h}_s^i+\mathbf{b}_i)+\mathbf{b}_s, \quad 1 \leq i \leq M\\
    \bar{h}_s^i = Concat[h_s^1, ..., h_s^{i-1}, h_s^{i+1},..., h_M],  \\
    h_{subpe}^{g_t} = \sum_{i=1}^M{\frac{exp(a_i)}{\sum_{i \in M}{exp(a_i)}} {h}_{s}^i}
\end{gather}
where $\mathbf{W}_{i}$, $\overline{\mathbf{W}}_{i}$, $\mathbf{W}_s$, $\mathbf{b}_i$ and $\mathbf{b}_s$ are trainable parameters. $a_i$ refers to the activation between any specific subset $s_i$ related to other subsets. $\bar{h}_s^i$ is the embedding of the all subsets within the group, except $h_s^i$. $h_{subpe}^{g_t}$ is the subset-granularity embedding for group $g_t$.

% employs an attention like MoSAN \cite{vinh2019interact} to capture the interaction and differences among subsets, improving weights for users with focused preferences. 

% \textbf{Sub-group Aggregation Module (SAM).}
% After getting the subgroup embedding $\hat{g}_{k,m,j}$ in Equation 3, the next step is to aggregate these sub-group embeddings to group embedding. In this module, we use another attention network to extract the interaction information between each sub-group, which overcomes the deficit that the model like AGREE \cite{AGREE} ignores the relatedness between users within a group.
% \begin{equation}
% P_{k,m,j}=W_2\sigma(W_{m,j}\hat{g}_{m,j} +W_{\bar{m},j}\hat{g}_{\bar{m},j}+b_1)+b_2,   m \in [1,M]\\
% \end{equation}
% where $\hat{g}_{m,j}$ is the sub-group embedding, and $\hat{g}_{\bar{m},j}$ is a set of sub-group embedding ${\hat{g}_1, \hat{g}_2, . . . , \hat{g}_M }$ exclude ${\hat{g}_m }$, and $\sigma$ is the $ReLU$ activation function. Through the softmax and weighted sum operations, we can get the group embedding which integrates the sub-groups' preferences and the user-user interaction:
% \begin{equation}
% g_{k,j}^{HAN} = \sum_{m=1}^M{\frac{exp(P_{k,m,j})}{\sum_{m \in M}{exp(P_{k,m,j})}} \hat{g}_{k,m,j}}
% \end{equation}

% \begin{equation}
% % a_{k,m,j} = ReLU(W_2(W_p\hat{g}_{k,p,j} + W_q\hat{g}_{k,q,j})+b_2),   p, q \in[1, M]\\
% % a_{k,m,j}=W_2\phi(W_mg_m +W_\bar{m}g_\bar{m}+b_1)+b_2,   m \in [1,M]\\
% \alpha_{k,m,j}=\frac{exp(P_{k,m,j})}{\sum_{m \in M}{exp(P_{k,m,j})}} \\
% \end{equation}

\subsection{Group Preference Extraction (GPE)}
% \textcolor{red}{Motivation: why we need to do it. How to achieve just display our equation.}
To better preserve the initial user expression and facilitate decision-making, we utilize the Group Preference Extraction (GPE) module to generate a distinct representation of group-level preferences,
\begin{equation}
    h_{gpe}^{g_t} = \sum_{k=1}^{\lvert gl \vert}{\frac{exp(\phi(\mathbf{W}_g(e(u_k)^T \cdot e(v_j))+\mathbf{b}_g))}{\sum_{k \in {\lvert gl \vert}}{exp(\phi(\mathbf{W}_g(e(u_k)^T \cdot e(v_j))+\mathbf{b}_g))}} e(u_k)}
\end{equation}
where $u_k \in g_t$ and $h_{gpe}^t$ refers to the group-granularity embedding of the group $g_t$. $\mathbf{W}_g$ and $\mathbf{b}_g$ are trainable parameters. 
\subsection{Superset Preference Extraction (SupPE)}

We use the Superset Preference Extraction (SupPE) module to gather external preference information at the superset-granularity, minimizing recommendation noise and gain a deeper understanding of group representation. This involves graph construction and external information extraction using SupPE. 
% We describe both stages later.

\textit{Graph Construction.}
Various groups may have members in common, leading to overlapping relationships among them. By utilizing these shared users, we can establish a graph structure that links different groups together. To construct the graph, we treat each group as a node, and if a user belongs to multiple groups, they act as an edge connecting the relevant nodes. Fig.~\ref{fig: model struct}(d) provides a visual demonstration of this concept.

\textit{External information Extraction.} By following the above approach, the graph can be constructed and external information regarding the behavioral preferences of users in other groups can be incorporated. This leads to a reduction in recommendation noise and a more comprehensive understanding of group representation. To obtain the superset-granularity embedding, follow these steps:
\begin{gather}
    h_{gl}^{l} = \phi(\mathbf{D}_g^{-\frac{1}{2}}\mathbf{A}_g\mathbf{D}_g^{-\frac{1}{2}}h_{gl}^{l-1}\mathbf{W}_g^l) \\
    h_{bt}^{l} = \phi(\mathbf{D}_b^{-\frac{1}{2}}\mathbf{A}_b\mathbf{D}_b^{-\frac{1}{2}}h_{bt}^{l-1}\mathbf{W}_b^l) \\
    h_{suppe}^{g_t} = Concat[h_{gl}^{l}, h_{bt}^{l}]
\end{gather}
where our approach involves $l$ layers and utilizes the initial group index ID embedding, denoted as $h_{gl}^{0}$, to aid in end-to-end training. We set $h_{bt}^{0}$ to be equal to $h_{subpe}^{g_t}$. The adjacency matrices for the global group and batch group are represented by $\mathbf{A}_g$ and $\mathbf{A}_b$, respectively, while $\mathbf{D}_g$ and $\mathbf{D}_b$ are the diagonal node degree matrices of $\mathbf{A}_g$ and $\mathbf{A}_b$. Global and batch groups are used to enhance embeddings and improve accuracy, while also increasing efficiency through regulation. $\mathbf{W}_g^l$ and $\mathbf{W}_b^l$ are trainable parameters.

% $l$ is the number of layers, $h_{gl}^{0}$ is the ID embedding of the initial group index, which can be help for end-to-end training. $h_{bt}^{0} = h_{subpe}^{g_t}$. $\mathbf{A}_g$ and $\mathbf{A}_b$ are the adjacency matrix for the whole group and batch group, respectively. $\mathbf{D}_g$ and $\mathbf{D}_b$ is the diagonal node degree matrix of adjacency matrix $\mathbf{A}_g$ and $\mathbf{A}_b$, respectively. $\mathbf{W}_g^l$ and $\mathbf{W}_b^l)$ are trainable parameters.
% \begin{equation}
% G^{GIE}=f(G,A)=\sigma(A_g \sigma(A_g(G^{inh} W^0_g))W^1_g)
% \end{equation}
% where $W^0$ and $W^1$ are trainable parameters. $\sigma$ is $ReLU$ activation function. $A_g$ is the adjacency matrix, and $G^{inh}$ is the inherent group embedding. It comes from group id feature and we can get it during training. After the GIE module, the model can learn all group-related features through the inherent group embedding during the training process. 

% \textbf{Local Group Information Extractor (LIE)}
% To bring the HAN embedding into our graph model, we introduce a local group information extractor that only uses group embedding in the same batch.
% The local group information extractor is proposed for dynamically aggregate user embeddings:

% \begin{equation}
% G^{LIE}_{batch}=f(G,A)=\sigma(A_l \sigma(A_l(G^{HAN}_{batch}W^0_l))W^1_l)
% \end{equation}
% where $A_l$ is the adjacency matrix in a batch. $G^{HAN}_{batch}$ is the embedding after sub-group aggregation in a batch. Different permutations and combinations will be chosen during the training process as part of our LIE's input.

\subsection{Fusion and Optimization}
%After previous work, we get 5 different kinds of embedding. How to effectively fusion the different kinds of embedding has a great impact in the model result. In our model, we use a self attention network to combine the 5 embedding vectors.
\textit{Fusion Layer.} The fusion layer is designed to dynamically incorporate preferences from the three levels of granularity embeddings (SubPE, GPE, and SupPE). Specifically, We employ a self-attention network to comprehensively uncover the latent preferences of users among various levels of granularity. This enables the network to flexibly adjust the weights between different embeddings, achieving a dynamic fusion of multi-level granularity preferences:
 % After the former layers, we get HAN, AN, GIE, LIE, and inherent embeddings. We employ the concatenate operation to get the combined vector $G_{con}$. Then, we use a one-layer network with the $ReLU$ activation function to convert the vector, and compute the attention score through the self-attention network:
\begin{gather}
h_{g_t} = Concat[h_{subpe}^{g_t}, h_{gpe}^{g_t}, h_{suppe}^{g_t}]\\
h_{fusion}^{{g_t}} = Softmax(\frac{h_{g_t} h_{g_t}^T}{\sqrt{d}}) \cdot h_{g_t}
\end{gather}
where $h_{g_t}$ is the concatenation embedding of the above three granularity embedding, and $d$ is the embedding dimension of $h_{g_t}$.

\textit{Prediction Layer.} Taking inspiration from AGREE\cite{AGREE}, our final prediction is derived through,
\begin{gather}
\hat{y}_{g_t} = \sigma(\mathbf{W}(Concat[h_{fusion}^{{g_t}},h_{fusion}^{{g_t}} \odot e(v_j) ,e(v_j)])+\mathbf{b}) 
\end{gather}
where $h_{fusion}^{{g_t}}$ is the group embedding after dynamic fusion. $\odot$ refers to Hadamard product and $\sigma$ is the sigmoid function. $\mathbf{W}$ and $\mathbf{b}$ are trainable parameters. 

\textit{Loss Function.} We update the entire network by minimizing the following loss $\mathcal{L}$ with the prediction $\hat{y}_{g_t}$ and the label $y_{g_t}$:
\begin{gather}
\mathcal{L}=\mathcal{L}_{triplet} + \lambda_1 \cdot\mathcal{L}_{point}\\
\mathcal{L}_{triplet} = \sum_{g_t\in{\mathcal{G}}}{[||\hat{y}_{g_t}-y^+_{g_t}||^2_2} - ||\hat{y}_{g_t}-y^-_{g_t}||^2_2 + \eta]_+ \\
\mathcal{L}_{point}=\sum_{g_t\in{\mathcal{G}}} {-[y_{g_t} log(\hat{y}_{g_t})}+(1-y_{g_t})log(1-\hat{y}_{g_t})]
\end{gather}
where $\lambda_1$ is the hyper-parameter, $y_{gt}^+$ and $y_{gt}^-$ are the prediction score of the item that has the same and different labels with the anchor, and $\eta$ is the margin.

\begin{table}[!ht]
\centering
\small
\vspace{-10pt}
\caption{Datasets Statistics. `ID', `Ml', `Mt-NYC', and `Mt-CA' stand for Industrial Dataset, Movielens-1M, Meetup-NYC, and Meetup-CA, respectively. }
% \caption{Datasets Statistics. `ID' =  Industrial Dataset, `Ml' = Movielens-1M, `Mt-NYC' = Meetup-NYC , and `Mt-CA' = Meetup-CA. }
\vspace{-10pt}
\label{tbl:dataset stastic}
\begin{tabular}{ ccccc}
 % \hline
 % \multicolumn{4}{|c|}{\bf{Dataset description}} \\
\toprule
 \bf{Dataset}& \bf{ID} &\bf{Ml} &\bf{Mt-NYC}&\bf{Mt-CA}\\
\midrule
Total Groups& 11,639&1,854&7,134&9,857\\

Total Users& 378,291&2,032&16,284&23,902\\

Total Items& 50,804,565&3,952&3,071&5,036\\

% Avg.Group Size&2.1&7.1&6.7\\
% Avg. Items/Group&32.8&1.0&1.0\\
% Avg. Items/User&177.9&3.1&2.8\\
% Avg. Users/Item&105.1&25.0&20.2\\
\bottomrule
\end{tabular}
\vspace{-10pt}

\end {table}

% \begin{table*}[!h]
% \centering
% \caption{Overall performance comparison on the Meetup-NYC and Meetup-CA datasets}
% \begin{tabular}{p{1.5cm}p{1.0cm}p{1.0cm}p{1.0cm}p{1.0cm}p{1.0cm}p{1.0cm}|p{1.0cm}p{1.0cm}p{1.0cm}p{1.0cm}p{1.0cm}p{1.0cm}}
%    \toprule
%    \bf{Model}&\multicolumn{6}{c|}{\bf{Meetup-NYC}}&\multicolumn{6}{c}{\bf{Meetup-CA}} \\
%    \bf{Metrics}  &P@5&P@10&R@5 &R@10&N@5&N@10 &P@5&P@10&R@5 &R@10&N@5&N@10\\
%    \midrule
%    NCF (AVG) &0.0296&0.0170&0.5999&0.6873&0.1135&0.1205&0.0317&0.0182&0.6327&0.7261&0.1257&0.1333\\
%    NCF (LM) &0.0296&0.0170&0.6027&0.6873&0.1136&0.1204&0.0318&0.0182&0.6307&0.7245&0.125&0.1327\\
%    NCF (MS) &0.0298&0.0170&0.5999&0.6867&0.1134&0.1204&0.0316&0.0182&0.6331&0.7253&0.1254&0.1329\\
%    NeuMF&0.0114&0.0084&0.2316&0.3411&0.0400&0.0487&0.0106&0.0095&0.2108&0.377&0.0328&0.0461\\
%    AGREE &0.0245&0.0154&0.4972&0.6226&0.0887&0.0987&0.0295&0.0179&0.5873&0.7152&0.1117&0.1221\\

%    MoSAN &0.0306&0.0176&0.6203&0.7132&0.1166&0.1240&0.0325&0.0185&0.6471&0.7363&0.1281&0.1354\\
%    GRADI &0.0369&0.0193&0.7482&0.7814&0.1714&0.1741&0.0366&0.0197&0.73&0.786&0.1735&0.1779\\
%    GAME &0.0378&0.0195&0.7679&0.7893&0.1808&0.1825&0.0369&0.0197&0.7378&0.7849&0.1787&0.1824\\
%    \bf{GHAN}&\bf{0.0382}&\bf{0.0196}&\bf{0.7744}&\bf{0.7924}&\bf{0.1834}&\bf{0.1848}&\bf{0.0375}&\bf{0.0199}&\bf{0.7479}&\bf{0.7939}&\bf{0.1789}&\bf{0.1826}\\
%    \bottomrule
% \end{tabular}
% \end{table*}

\begin{table*}[!h]
\centering
\small
\label{tbl:sota}
\caption{Performance Comparison of Various Methods. `SC'=Spectral Clustering, `AC'= Agglomerative Clustering.}
\vspace{-10pt}
\begin{tabular}{p{1.8cm}|p{0.9cm}|p{0.9cm}|p{0.9cm}|p{0.9cm}|p{0.9cm}|p{0.9cm}|p{0.9cm}|p{0.9cm}|p{0.9cm}|p{0.9cm}|p{0.9cm}|p{0.9cm}   }

 \toprule
  \multirow{2}{*}{\bf{Model}}&\multicolumn{4}{c|}{\bf{Movielens}}&\multicolumn{4}{|c|}{\bf{Meetup-NYC}}&\multicolumn{4}{|c}{\bf{Meetup-CA}} \\
   % \hline
   \cline{2-13}
 ~ & HR@5&HR@10&N@5&N@10 &HR@5 &HR@10&N@5&N@10 &HR@5 &HR@10&N@5&N@10\\
 \hline

NCF (AVG) &0.3891&0.5886&0.3899&0.4561&0.5999&0.6873&0.1135&0.1205&0.6327&0.7261&0.1257&0.1333\\
NCF (LM) &0.3936&0.5895&0.3926&0.4571&0.6027&0.6873&0.1136&0.1204&0.6307&0.7245&0.1250&0.1327\\
NCF (MS) &0.3910&0.5869&0.3910&0.4558&0.5999&0.6867&0.1134&0.1204&0.6331&0.7253&0.1254&0.1329\\

\hline

NeuMF&0.4057&0.5940&0.4110&0.4702&0.2316&0.3411&0.0400&0.0487&0.2108&0.3770&0.0328&0.0461\\
AGREE &0.4224&0.6171&0.4121&0.4761&0.4972&0.6226&0.0887&0.0987&0.5873&0.7152&0.1117&0.1221\\

MoSAN &0.4032&0.6000&0.4012&0.4659&0.6203&0.7132&0.1166&0.1240&0.6471&0.7363&0.1281&0.1354\\
GRADI &0.3503&0.5292&0.3468&0.4064&0.7482&0.7814&0.1714&0.1741&0.7300&0.7860&0.1735&0.1779\\
GAME &0.3858&0.5554&0.3909&0.4447&0.7679&0.7893&0.1808&0.1825&0.7378&0.7849&0.1787&0.1824\\
\hline \hline 

% only AN &0.3967&0.5575&0.3977&0.4483&0.7498&0.7732&0.1761&0.1780&0.7215&0.7670&0.1745&0.1782\\

% only HAN &0.4167&0.5987&0.4232&0.4805&0.7706&0.7864&0.1822&0.1834&0.7456&0.7936&0.1793&0.1831\\

 w/o SubPE&0.3939&0.5563&0.3865&0.4397&0.7551&0.7765&0.1776&0.1793&0.7357&0.7866&0.1764&0.1805\\
 
 w/o GPE&0.4131&0.5951&0.4124&0.4722&0.7710&0.7888&0.1826&0.1840&0.7432&0.7982&0.1761&0.1805\\

 w/o SupPE&0.4413&0.6298&0.4401&0.5011&0.7626&0.7822&0.1798&0.1814&0.7236&0.7622&0.1748&0.1779\\
 
%  w/o IH&0.4131&0.5951&0.4124&0.4722&0.7720&0.7879&0.1832&0.1845&0.7400&0.7916&0.1759&0.1801\\

% w/o HAN &0.4194&0.6053&0.4148&0.4765&0.7724&0.7881&0.1831&0.1843&0.7399&0.7883&0.1758&0.1797\\

% w/o AN&0.4232&0.6101&0.4187&0.4803&0.7735&0.7884&0.1830&0.1842&0.7403&0.7935&0.1758&0.1801\\
% w/o GIE&0.4033&0.5875&0.4027&0.4627&0.7708&0.7868&0.1830&0.1843&0.7282&0.7773&0.1736&0.1776\\
% w/o LIE&0.4460&0.6252&0.4430&0.4998&0.7703&0.7872&0.1828&0.1841&0.7461&0.7981&0.1791&0.1833\\
\hline \hline 
MGAM w/ SC &0.4348&0.6241&0.4289&0.4913&0.7580&0.7801&0.1783&0.1801&0.7410&0.7898&0.1782&0.1821\\

MGAM w/ AC &0.4112&0.6014&0.4028&0.4673&0.7536&0.7784&0.1781&0.1800&0.7391&0.7869&0.1772&0.1810\\
\hline \hline 

\bf{MGAM}&\bf{0.4458}&\bf{0.6335}&\bf{0.4403}&\bf{0.5015}&\bf{0.7741}&\bf{0.7938}&\bf{0.1835}&\bf{0.1850}&\bf{0.7460}&\bf{0.7930}&\bf{0.1790}&\bf{0.1827}\\

\bottomrule
\end{tabular}
\vspace{-10pt}
\label{tbl:sota}
\end{table*}

\section{Experiments}

\subsection{Experimental Settings}

\subsubsection{Datasets \& Metrics}Our experiments were conducted on four datasets, namely, Movielens-1M, Meetup-NYC, Meetup-CA, which are extensively used in real-world scenarios, and an industrial dataset. Details can be found in Table~\ref{tbl:dataset stastic}. For Movilens-1M, the way we generate groups is line with the approach in \cite{baltrunas2010group}. Events in Meetup datasets are groups with attendees as members and the venue as the item. The Industrial dataset collected from Ele.me, a major online food ordering platform in China, includes seven consecutive days of training data and one day of testing data. The detailed implementation of our industrial dataset is outlined in Section~\ref{sec:industrial implementation}. In terms of metrics, we adopt Hit Ratio and Normalized Discounted Cumulative Gain at top-\textit{K} recommendations, known as \textit{HR@K} and \textit{NDCG@K}, respectively, are widely used in recommendation systems. Higher \textit{HR@K} and \textit{NDCG@K} indicate better
performance. In analyzing industrial datasets, we employ the Area Under the Curve (AUC) metric in offline experiments and utilize Click-Through Rate (CTR) to measure online performance. These metrics are commonly employed in industrial systems.
% \subsubsection{\bf{MovieLens 1M}}

% The Movielens 1m dataset contains 1 million moving ratings from over 6,000 users on approximately 4,000 movies. Moreover, the dataset is commonly used in recommendation tasks. Following the approach in \cite{baltrunas2010group}, we randomly generate some groups from all of the users. If every member gives four stars or above to a movie, we assume that the group adopts this movie. In this scenario, we simulate the group formed between unrelated people.

% \subsubsection{\textbf{Meetup-NYC, CA}}

% Meetup datasets includes two sub-datasets, Meetup-NYC and Meetup-CA, containing the events held in New York City and California, respectively. Following previous works' definition, we consider an event as a group, and the user participating in the event as a member, and a venue as an item. This task aims to recommend a venue for a given group to host an event.

% \subsection{Evaluation metrics}

% To ensure fair evaluations, we apply these metrics to all methods using the same test data.

% \subsection{Comparison Method \& Details}

% For Meetup-NYC and Meetup-CA datasets, we applied dropout \cite{srivastava2014dropout} with keep probabilities of 0.5 and 0.7, respectively, before every fully-connected layer. The batch size is set to 256 for both datasets, and the embedding dimensions are set to 32 and 64 for Meetup-NYC and Meetup-CA, respectively.
\subsubsection{Comparison Method \& Running Settings}To verify the effectiveness of MGAM, we select three memory-based methods (NCF AVG \cite{he2017neural}, NCF LM \cite{he2017neural}, NCF MS \cite{he2017neural}) and five attention-based methods (NeuMF \cite{neumf}, AGREE \cite{AGREE}, MoSAN \cite{vinh2019interact}, GRADI \cite{2019GRADI}, GAME \cite{GAME}) for comparison. All methods are implemented with Tensorflow1.14. Adam optimizer \cite{kingma2014adam} with a learning rate of 0.001 optimizes the objective function.  The cluster method is chosen in Section.~\ref{sec:subset generation} is K-Means.  $M$ was grid searched to 3 and 5 for Movielens-1M and Meetup, respectively. The layers of $l$ is set to 2, $\lambda_1$ is 0.5 and the margin $\eta$ is set to 1. All the experiments in our paper have been repeated 10 times, with the final output reflecting the average scores.

\begin{figure}[tbp]
  \centering
  % \vspace{-10pt}
  \includegraphics[width=1.\linewidth]{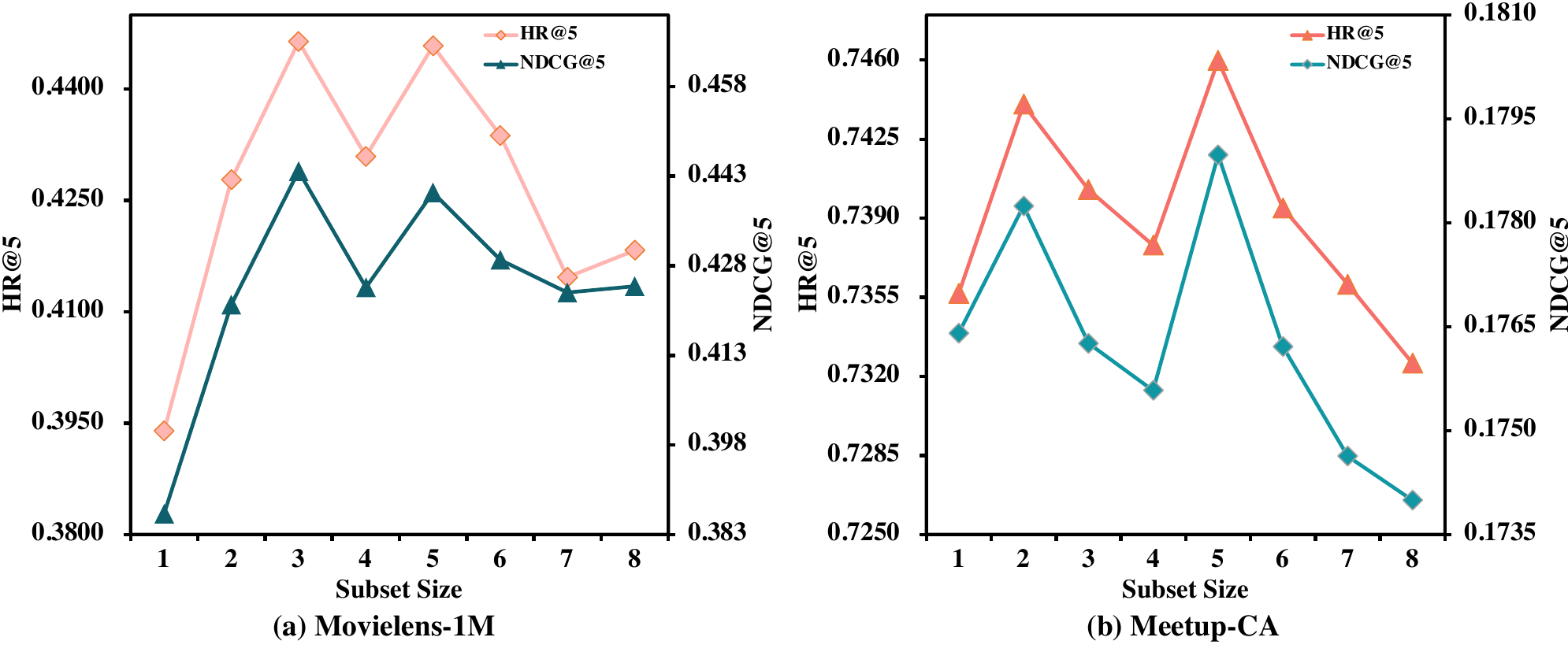}
  \vspace{-10pt}
  \caption{Performance of different subsets' size on Movielens-1M dataset (a), and Meetup-CA dataset (b).}
  \vspace{-15pt}
  \label{group_size}
\end{figure}

\subsection{Experiments Results}
%We compared our model with several state of the art baseline in our experiments. the result shows that our model has better performance than agree and mosan. In the following chapter, we will disassemble the our and analysis the importance of different part.

% We repeat these experiments 10 times and report the average scores as the final performances. 
\subsubsection{Performance Comparison.}
The results of our experiments are listed in Table~\ref{tbl:sota}. We can identify that the majority of model-based methods outperform memory-based methods, which is a common trend. Our MGAM has attained a state-of-the-art performance in three datasets across various metrics.

% \textcolor{red}{\# todo, determined by the adding experiments.}

% The p-values are lower than 0.0001.???? 

% According to the comparison of the results, we have some observations. 

% (1) In Meetup-NYC and Meetup-CA, MoSAN shows its strength in considering the attention between users. Furthermore, GAME considers the influences from the counterpart views when modeling users and items and performs better than GRADI and MoSAN. 

% (2) GHAN significantly outperforms other baselines, including the traditional neural-based methods ( NCF, NeuMF ), the neural attention networks ( AGREE, MoSAN ), and the hybrid model ( GRADI, GAME ),  achieving state-of-the-art performance on both datasets. Specifically, considering multiple views of group information, our hybrid model shows an equilibrium performance in different datasets.

\subsubsection{Ablation Study.}
% \subsection{Model analysis}
%In this part, we will do some experiments and analysis to investigate the influence of  components (including model structure and group embedding), group sizes and group diversity in our model.
% The result of the ablation study for each module is illustrated in Table \ref{tbl:sota}. It is obvious that removing any of the modules reduces GHAN's effectiveness. demonstrating their positive impacts on group recommendation scenarios.
Table \ref{tbl:sota} also presents the outcome of the ablation study conducted on each module, which clearly shows that MGAM's performance is negatively impacted by removing any of the modules. This ablation study demonstrates that each granularity module (denoted as w/o SubPE, w/o GPE, and w/o SupPE) plays a distinct role in capturing the users' latent preferences. Especially, removing SubPE significantly decreases performance compared to other modules, indicating the superiority introduced by the subset granularity.

\subsubsection{Impact of Subsets' Number}
Fig.~\ref{group_size} presents the results from hyper-parameter tuning for the number of subsets within a group. Results were only reported for Movielens-1M and Meetup-CA due to similar results on Meetup-NYC. Higher $M$ indicates more subsets in a group and fewer members in a subset. 
The optimal subset sizes for the two datasets are 3 and 5, as depicted in Fig.~\ref{group_size}(a) and Fig.~\ref{group_size}(b), respectively.

% Fig.~\ref{group_size}(a) shows fluctuating pattern, while Fig.~\ref{group_size}(b) shows initial rise and decline with optimal point at size 3 and 5 for respective datasets.

\subsubsection{Analysis of Clustering Methods}
Table~\ref{tbl:sota} also shows variation studies with different clustering methods, including K-Means (MGAM), Spectral Clustering (MGAM w/ SC), and Agglomerative Clustering (MGAM w/ AC). Our findings suggest that our approach is better suited to be used with K-Means compared to Spectral Clustering and Agglomerative Clustering. This is likely due to K-Means clustering having several advantages over Agglomerative Clustering and Spectral Clustering, including scalability, simplicity, and convergence guarantees.

\begin{table}[tbp]
    \centering
    \small
    % \vspace{-10pt}
    \begin{tabular}{ ccc}
 \toprule
 \textbf{Model} & \textbf{Offline AUC} & \textbf{Online Improvement (CTR)} \\
 \midrule
AGREE & 0.6570  & - \\
\textbf{MGAM}  & 0.6880  & +1.2\% \\
\bottomrule

\hline
\end{tabular}
    \caption{Offline and online performance on industrial dataset.}
    \label{tab:industrial_dataset}
    \vspace{-25pt}
\end{table}

\subsubsection{Industrial Implementation and Online Performance}
\label{sec:industrial implementation}
% \textcolor{red}{todo hehe }In practice, $i.e.$, online food ordering service, we implement the area grid as area splition. In specail area grid, those people has buying history in pass 7 days would consist of a group. Since they share the similar area preference. For these group users, we make group recommendation for product collection, which is the stage right just before the product recall stage in industrial recommendation. In this case, we can efficiently improve the product selecting without consuming lots of  human labors and enhance the efficiency of this stage, thereby improving the final personal recommendation result. 
In online food ordering services, an area grid is utilized as a means of area division. Specifically, for a designated area grid, individuals with purchasing histories within the past seven days are grouped since they exhibit similar area preferences. For such groups of users, we can employ group recommendation techniques to suggest product collections just prior to the product recall stage in industrial recommendation. By doing so, we can significantly improve product selection without the need for excessive human labor and enhance the efficiency of this stage. Ultimately, this approach can lead to improved personalized recommendation outcomes. To validate the performance of our proposed model, we conducted methods comparison in our industrial dataset. Table~\ref{tab:industrial_dataset} shows that our MGAM significantly outperformed the base model (AGREE). Our online experiments also confirmed this, as we deployed MGAM and compared it with the AGREE. The CTR from the 5-day experiment consistently outperformed the base model, with an average improvement of 1.2\%.

% To validate the performance of our proposed model, we conduct methods comparison in our Industrial dataset. Observed from Table~\ref{tab:industrial_dataset}, we found our MGAM have improved from AGREE large marginally, with also been proved by our online experiments. We deployed MGAM online and compared it with the base model (AGREE \cite{AGREE}). The CTR predictions of the 5-days experiment consistently outperformed the baseline, with and average improvement of 1.2\%.

% \textcolor{red}{tell the detail how to define a group for our industrial implementation, which is online food ordering service. Then, why we select agree as our base model. Because of its running efficiency? Then online performance comparison. Which stage can our model implement} We deployed MGAM online and compared it with the base model (AGREE \cite{AGREE}). The CTR predictions of the 5-days experiment consistently outperformed the baseline, with and average improvement of 1.2\%. 

\section{Conclusions}

In this paper, we propose a Multi-Granularity Attention Model (MGAM) to uncover the latent preferences of users in group recommendation tasks, further improving the preference extraction and aggregation ability of the model. Specifically, we propose the Subset Preference Extraction module, the Group Preference Extraction module, and the Superset Extraction Preference module from 3 different granularities to uncover group members' latent preferences and further mitigate recommendation noise. Extensive offline experiments and online performance demonstrate the effectiveness and efficiency of MGAM.

% , especially for sparse user activity groups. We will extend the model to introduce the time and position information to enhance the recommendation performance in future work. Moreover, it is hard to build a group graph if users only exist in one group in some situations. We will also improve this defect in the follow-up work.

\bibliographystyle{ACM-Reference-Format}
\bibliography{ref}

\end{document}